\documentclass[10pt]{blois07}
\usepackage{graphicx}
\usepackage{cite,./mcite}
\usepackage{pgf,pgfarrows,pgfnodes,pgfautomata,pgfheaps,pgfshade}
\usepackage{pstricks,pst-node,pst-text,pst-3d}
\usepackage{amssymb}
\usepackage{xspace}
\usepackage{amsmath}
\usepackage{picinpar}
\usepackage{subfigure}

\title{On the measurement of the proton-air cross section \\using cosmic ray
  data}
\author{Ralf~Ulrich$^{\rm 1,}$\footnote{$\;\;$ralf.ulrich@kit.edu}$\;\;$,  Johannes~Bl{\"u}mer$^{\rm 1,2}$, Ralph~Engel$^{\rm 1}$, Fabian~Sch{\"u}ssler$^{\rm 1}$, Michael~Unger$^{\rm 1}$}
\institute{$^{\rm 1}$Institut f\"ur Kernphysik, Forschungszentrum Karlsruhe,
  Karlsruhe, Germany\\$^{\rm 2}$Institut f\"ur Experimentelle Kernphysik,
  Universit{\"a}t Karlsruhe, Karlsruhe, Germany}

\begin{document}
  \maketitle
  \begin{abstract}
    Cosmic ray data may allow the determination of the proton-air
    cross section at ultra-high energy. For example, the distribution
    of the first interaction point in air showers reflects
    the particle production cross section. As it is not possible to
    observe the point of the first interaction $X_{\rm 1}$ of a cosmic
    ray primary particle directly, other air shower observables must be linked
    to $X_{\rm 1}$.  This introduces an inherent dependence of the
    derived cross section on the general understanding and modeling of
    air showers and, therfore, on the hadronic interaction model used
    for the Monte Carlo simulation. We quantify the
    uncertainties arising from the model dependence by varying some
    characteristic features of high-energy hadron production.
  \end{abstract}
  
  \section{Introduction}

  The natural beam of cosmic ray particles extends to energies far
  beyond the reach of any earth-based accelerator. Therefore cosmic
  ray data provides an unique opportunity to study interactions at
  extreme energies. Unfortunately, the cosmic ray flux is extremely
  small making direct measurements of the particles and their interactions impossible above
  $\sim$ 100 TeV. One is forced to rely on indirect measurements such
  as extensive air shower studies, where interpretation of the data is
  very difficult.
  
\begin{figure}[t!]
    \centering
    \includegraphics[width=.8\linewidth]{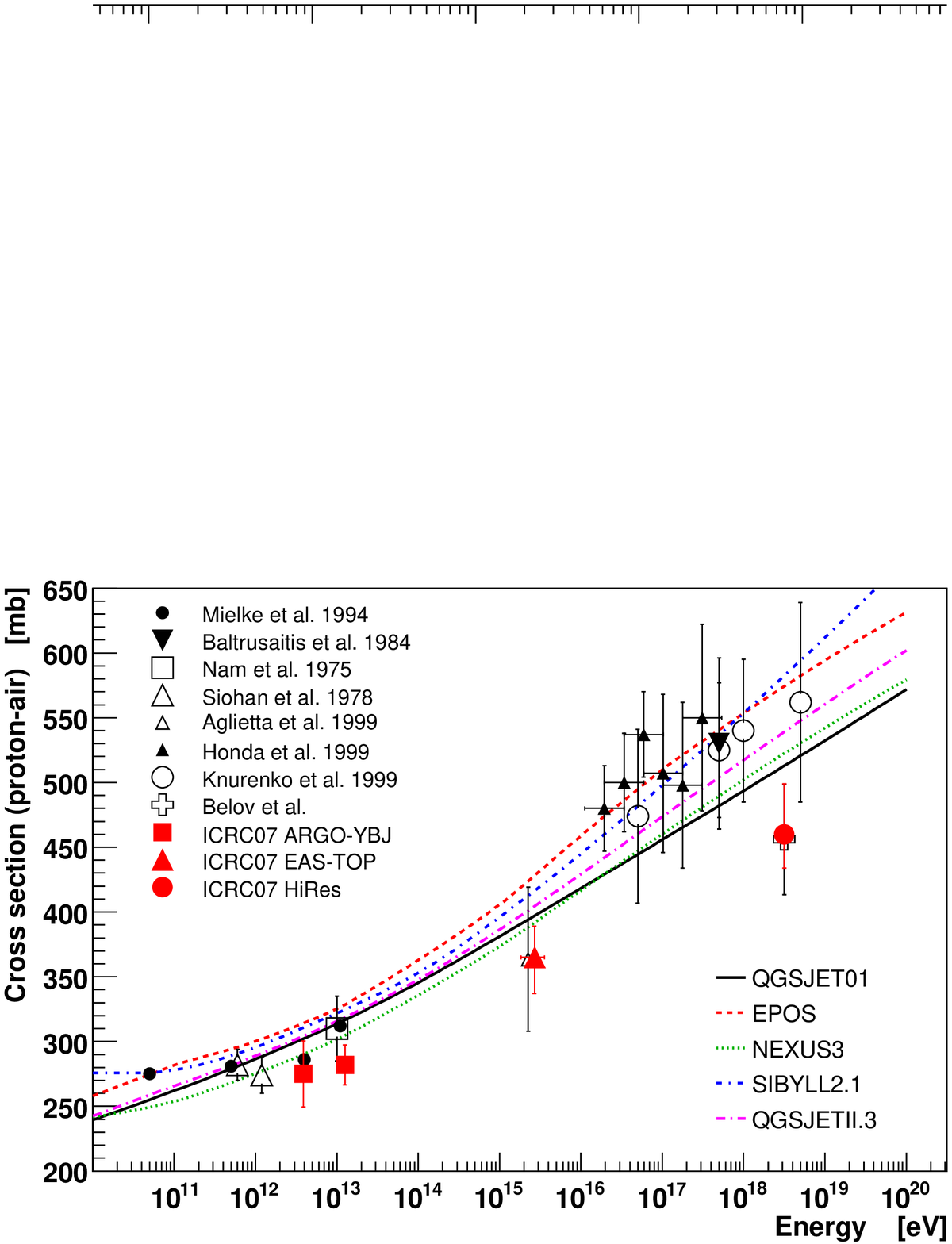}
    \caption{
Current data of proton-air production cross section 
      measurements   
  \protect\cite{Grigorov:1965aa,Yodh:1972fv,Nam:1975aa,Siohan:1978zk,Mielke94,%
   Baltrusaitis:1984ka,%
   Honda:1993kv,Aglietta:1999bd,Hara:1983pa,Knurenko:1999cr,%
   Belov:2006mb}
   and model predictions 
   \protect\cite{Ranft:1994fd,Pierog:2006qv,Drescher:2000ha,Kalmykov:1997te,%
     Ostapchenko:2005nj,Engel:1999db,Fletcher:1994bd}.
}
    \label{f:data}
  \end{figure}

  In this contribution we will briefly discuss different methods of measuring
  the proton-air cross section, focusing on methods that are based on
  extensive air shower (EAS) data. Figure~\ref{f:data} shows a
  compilation of proton-air cross section measurements and predictions
  of hadronic interaction models currently used in cosmic ray studies
  \cite{Grigorov:1965aa,Yodh:1972fv,Nam:1975aa,Siohan:1978zk,Mielke94,%
   Baltrusaitis:1984ka,%
   Honda:1993kv,Aglietta:1999bd,Hara:1983pa,Knurenko:1999cr,%
   Belov:2006mb}


  \section{Methods of  cross section measurements using cosmic ray data}

  \subsection{Primary cosmic ray proton flux}

  Already in the 60's first estimates of the proton-air cross section
  $\sigma_{\rm p-air}$ were made using cosmic ray
  data \cite{Grigorov:1965aa}. These early measurements are relying on two
  independent observations of the flux of primary cosmic ray protons
  after different amounts of traversed atmospheric matter. Firstly the
  primary proton flux $\Phi(X_{\rm top})$ is measured at the top of
  the atmosphere with a satellite or at least very high up in the
  atmosphere on a balloon at $X_{\rm top}=0-5$~gcm$^{-2}$. The second
  flux $\Phi(X_{\rm bottom})$ is measured with a ground based
  calorimeter at $X_{\rm bottom}=600-1000$~gcm$^{-2}$, preferentially
  at high altitude and using efficient veto detectors to select
  unaccompanied hadrons. The effective attenuation length can then be calculated
  straightforwardly from
  \begin{equation}
    \lambda_{\rm prod} = (X_{\rm bottom}-X_{\rm top}) / \log(\Phi_{\rm
      top}/\Phi_{\rm bottom}) . 
  \end{equation}
  As it is impossible to veto all hadronic interactions along the
  cosmic ray passage through the atmosphere, this attenuation length
  can only be used to obtain a lower bound to the high energy particle
  production cross section
  \begin{equation}
  \sigma_{\rm p-air} \ge \frac{\langle m \rangle}{\lambda_{\rm prod}} ,
  \end{equation}
  where $\langle m \rangle$ is the mean mass of air.
  The method is limited to proton energies lower than $\sim$TeV, since no sufficiently precise 
  satellite or balloon borne data is available above this energy. By design the
  unaccompanied hadron flux is only sensitive to the particle 
  production cross section, since primary protons with interactions without
  particle production cannot be separated from protons without any
  interaction. 

  \subsection{Extensive air showers}

  In order to measure $\sigma_{\rm p-air}$ at even higher energies it is
  necessary to rely on EAS data
  \cite{Baltrusaitis:1984ka,Honda:1993kv,Aglietta:1999bd,Hara:1983pa,Knurenko:1999cr,%
    Belov:2006mb}. The
  characteristics of the first few extremely high energy hadronic interactions
  during the startup of an EAS are paramount for the resulting air
  shower. Therefore it should be possible to relate EAS observations like the
  shower maximum $X_{\rm max}$, or the total number of electrons 
  $\left.N_{\rm e}(X)\right|_{X=X_{\rm obs}}=N_{\rm e}^{\rm rec}$
  and muons 
  $\left.N_{\mu}(X)\right|_{X=X_{\rm obs}}=N_{\mu}^{\rm rec}$
  at a certain observation depth $X_{\rm obs}$, to the depth of the first
  interaction point and the characteristics of the high energy hadronic
  interactions.\vspace*{.1cm}\\ 
  {\bf Ground based observations}

\noindent
  In case of ground based extensive air shower arrays, the frequency
  of observing EAS of the same energy at a given stage of their development is used for the cross
  section measurement. By selecting EAS of the same energy but
  different directions, the point of the first interaction has to vary
  with the angle to observe the EAS at the same development stage. The
  selection of showers of constant energy and stage depends on the particular
  detector setup, but the typical requirement is 
  $\left( N_{\rm e}^{\rm rec},\;N_{\mu}^{\rm rec}\right)=\rm{const}$ at observation level.

  \begin{figure}[t!]
    \centering
    \includegraphics[width=.6\linewidth]{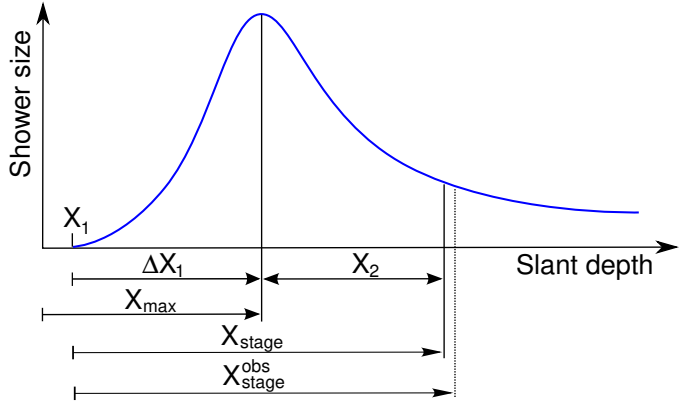}
    \caption{Definition of variables to characterize EAS longitudinal profiles.}
    \label{f:ShowerStages}
  \end{figure}  

  With the naming conventions given in Fig.~\ref{f:ShowerStages}, the
  probability of observing a shower of a given energy $E_0$ and shower stage
  at the zenith angle $\theta$ can be written as 
%
  \begin{eqnarray}
    \frac{1}{N}\left.\frac{dN}{d\cos\theta}\right|_{N_{\rm e}^{\rm rec}, N_{\mu}^{\rm rec}}
    &=& \int dX_1 \int d\Delta X_1 \int d\Delta X_2 \;\;
    \frac{e^{-X_{\rm 1}/\lambda_{\rm int}}}{\lambda_{\rm int}} \nonumber\\
    &&\times\; P_1(\Delta X_1) \times  P_2(\Delta X_2) \nonumber\\
    &&\times\; P_{\rm res}(X^{\rm rec}_{\rm stage}, X_1+\Delta X_1+\Delta X_2). 
    \label{eqn::frequency}
  \end{eqnarray} 
  Here $X_{\rm stage}$ defines the distance between the first
  interaction point and the depth at which the shower reaches a given
  number of muons and electrons as defined by the selection
  criteria. The experimentally inferred shower stage at observation
  level $X^{\rm rec}_{\rm stage}$ does, in general, not coincide with
  the true stage due to the limited detector and shower reconstruction
  resolution. This effect is accounted for by the factor $P_{\rm
    res}$. The functions $P_1$ and $P_2$ describe the shower-to-shower
  fluctuations. The probability of a shower having its maximum at
  $X_{\rm max} = X_1 + \Delta X_1$ is expressed by $P_1$. The
  probability $P_2$ is defined correspondingly with $X_{\rm stage} =
  \Delta X_1 + \Delta X_2$.

  In cross section analyses, Eq.~(\ref{eqn::frequency}) is
  approximated by an exponential function of $\sec\theta$. 
  Assuming that the integration of (\ref{eqn::frequency}) over the
  distributions $P_1$, $P_2$, and $P_{\rm res}$ does not yield any generally
  non-exponential tail at large $\sec\theta$, it can be written as 
  \begin{equation}
    \label{eqn:LambdaSD}
    \frac{1}{N}\left.\frac{dN}{d\cos\theta}\right|_{N_{\rm e}^{\rm rec}, N_{\mu}^{\rm rec}} \;\propto\; e^{-X_{\rm obs}/\Lambda_{\rm obs}^{\rm S}} \;\propto\;
    e^{-\sec\theta/\Lambda_{\rm obs}^{\rm S}}.
  \end{equation}
  However, the slope parameter $\Lambda_{\rm obs}^{\rm S}$ does
  not coincide with the interaction length $\lambda_{\rm int}$ due
  to non-Gaussian fluctuations and a possible angle-dependent
  experimental resolution. Therefore the
  measured attenuation length can be written as
  \begin{equation}
    \label{eqn:FrequencyLambda}
    \Lambda_{\rm obs}^{\rm S} = \lambda_{\rm int} \cdot k_{\Delta X_{\rm 1}} \cdot k_{\Delta X_{\rm 2}} \cdot k_{\rm resolution}^{\rm S} = \lambda_{\rm int}
    \cdot k_{\rm S}.
  \end{equation} 
  The $k$-factors $k_{\Delta X_{\rm 1}}$, $k_{\Delta X_{\rm 2}}$ and $k_{\rm
    resolution}^{\rm S}$ parametrize the contributions to
  $\Lambda_{\rm obs}^{\rm S}$ from the corresponding
  integrations. However, these integrations are
  difficult to perform separately and the individual $k$-factors are not 
  known in most analyses (for a partial exception, see
  \cite{Knurenko:1999cr}).\vspace*{.1cm}\\ 
  {\bf Observations of the shower maximum $\mathbf{X_{\rm max}}$}
  \begin{figure}[t!]
    \centering
    \includegraphics[width=.8\linewidth]{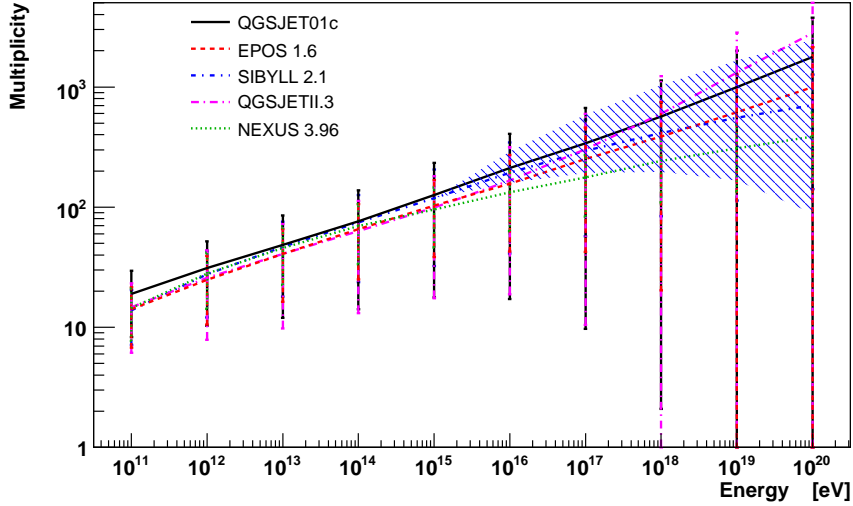}
    \caption{Model predictions for the secondary particle multiplicity in
    high energy hadronic interactions. The
    lines denote the mean while the error bars indicate the RMS of the
    distributions. The shaded area is the range for SIBYLL using $0.3\ge
    f_{\rm 10EeV}\ge 3$.}   
    \label{f:mult}
  \end{figure}

  \noindent
  Observing the position of the shower maximum directly allows one to
  simplify (\ref{eqn::frequency}) by removing the term due to the
  shower development after the shower maximum $P_2$. Also
  the detector resolution $P_{\rm res}$ is
  much better under control for $X_{\rm max}$ and can be well approximated
  by a Gaussian distribution. The resulting distribution is
  \begin{eqnarray}
    \label{eqn::XmaxTail}
    P(X_{\rm max}^{\rm rec}) &=&  \int dX_{\rm 1}
    \int d\Delta X_{\rm 1} \;\; 
    \frac{e^{-X_{\rm 1}/\lambda_{\rm int}}}{\lambda_{\rm int}}
    \times\; P_1(\Delta X_1) \times\; P_{res}(X_{\rm max}^{\rm rec}-X_{\rm max}),
  \end{eqnarray} 
  with $X_{\rm 1}+\Delta X_{\rm 1}=X_{\rm max}$.
  In analogy to Eq.~(\ref{eqn:LambdaSD}) only the tail of $P(X_{\rm max}^{\rm rec})$ at
  large $X_{\rm max}^{\rm rec}$ is approximated by an exponential distribution 
  \begin{equation}
    P(X_{\rm max}^{\rm rec}) \;\propto\; e^{-X_{\rm max}^{\rm rec}/\Lambda_{\rm obs}^{\rm X}},
  \end{equation}
  whereas the exponential slope $\Lambda_{\rm obs}$ can be deduced from the
  convolution integral (\ref{eqn::XmaxTail}) as
  \begin{equation}
    \label{eqn:XmaxTailLambda}
    \Lambda_{\rm obs}^{\rm X} = \lambda_{\rm int} \cdot k_{\Delta X_{\rm 1}} 
  \cdot k_{\rm resolution}^{\rm X} = \lambda_{\rm int} \cdot k_{\rm X}.
  \end{equation}
  Again $k_{\Delta X_{\rm 1}}$ and $k_{\rm resolution}^{\rm X}$
  are the contributions to $\Lambda_{\rm obs}^{\rm X}$ from the corresponding
  integrations of~(\ref{eqn::XmaxTail}). \\
  It was also recognized that (\ref{eqn::XmaxTail}) can be unfolded
  directly to retrieve the original $X_{\rm 1}$-distribution, if the
  $\Delta X_{\rm 1}$-distribution is previously inferred by Monte-Carlo simulations
  \cite{Belov:2006mb}. Recently this triggered some discussion about the general
  shape and model dependence of the $\Delta X_{\rm 1}$-distribution
  \cite{Ulrich:2006hb}. This directly implies a corresponding model dependence of the
  $k_{\Delta X_{\rm 1}}$-factors. 
  
  \section{Impact of high energy interaction model characteristics on air
    shower development}
  \label{sec:HEimpact}
  \begin{figure}[t!]
    \includegraphics[width=.33\linewidth]{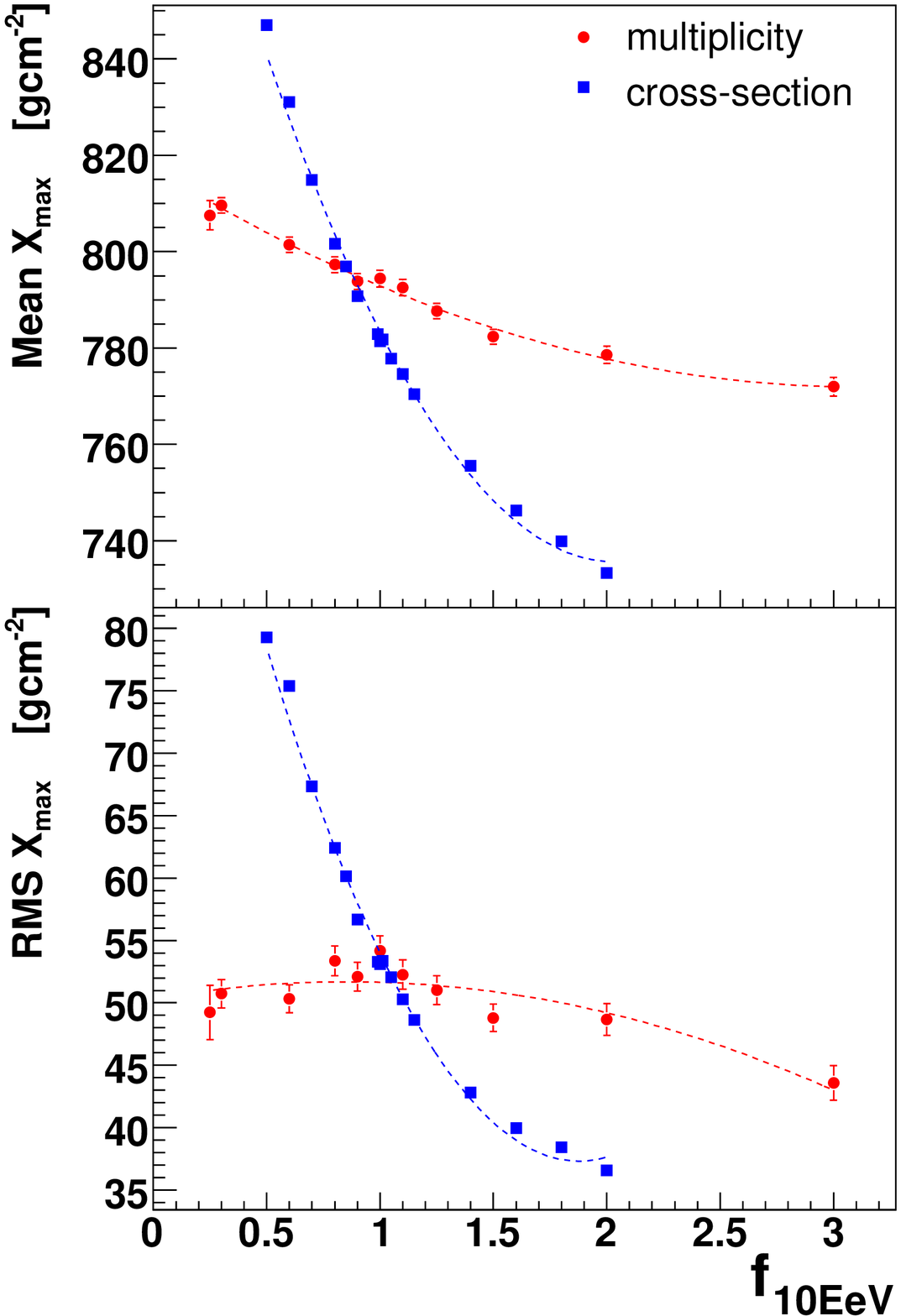}~
    \includegraphics[width=.33\linewidth]{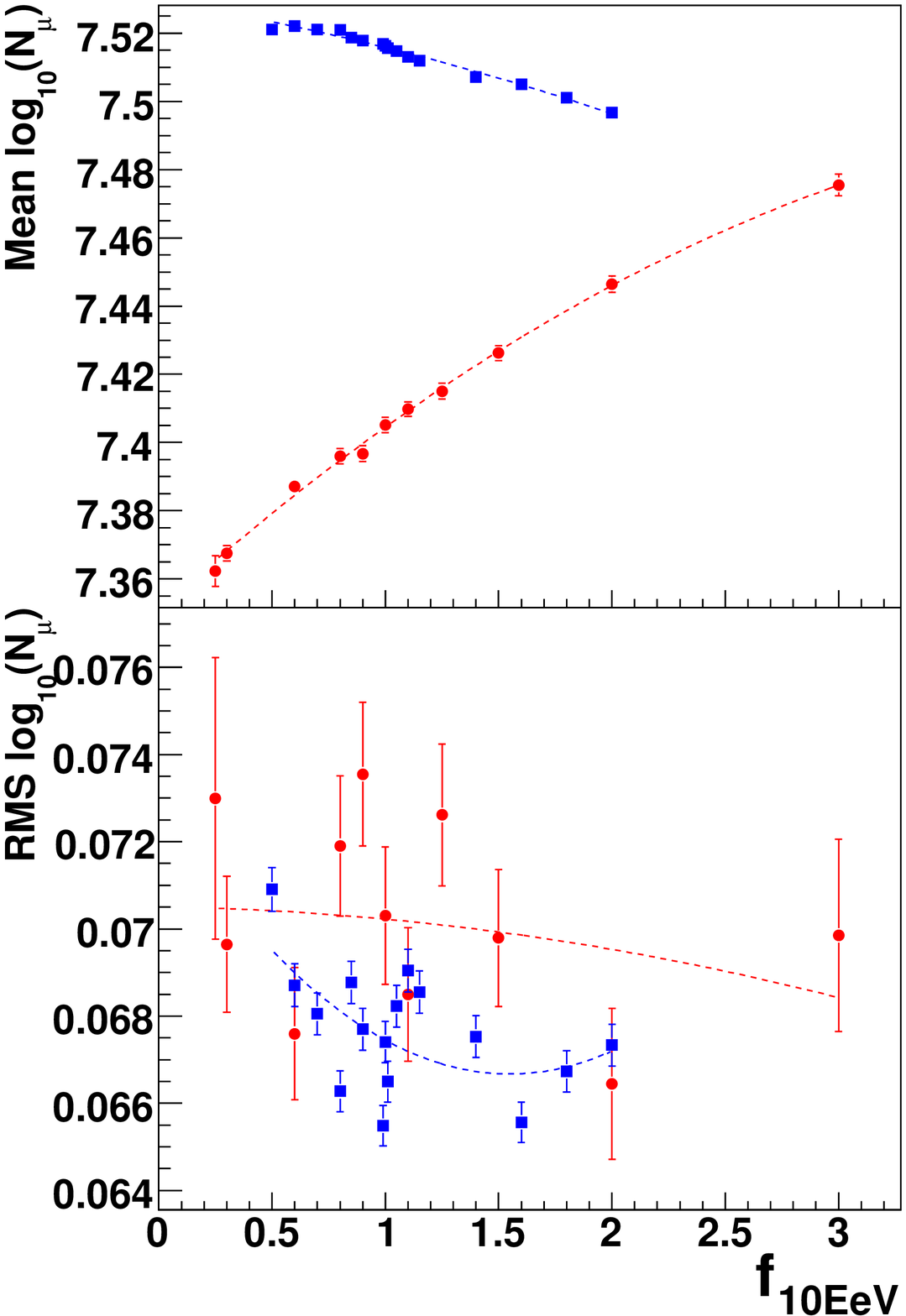}~
    \includegraphics[width=.33\linewidth]{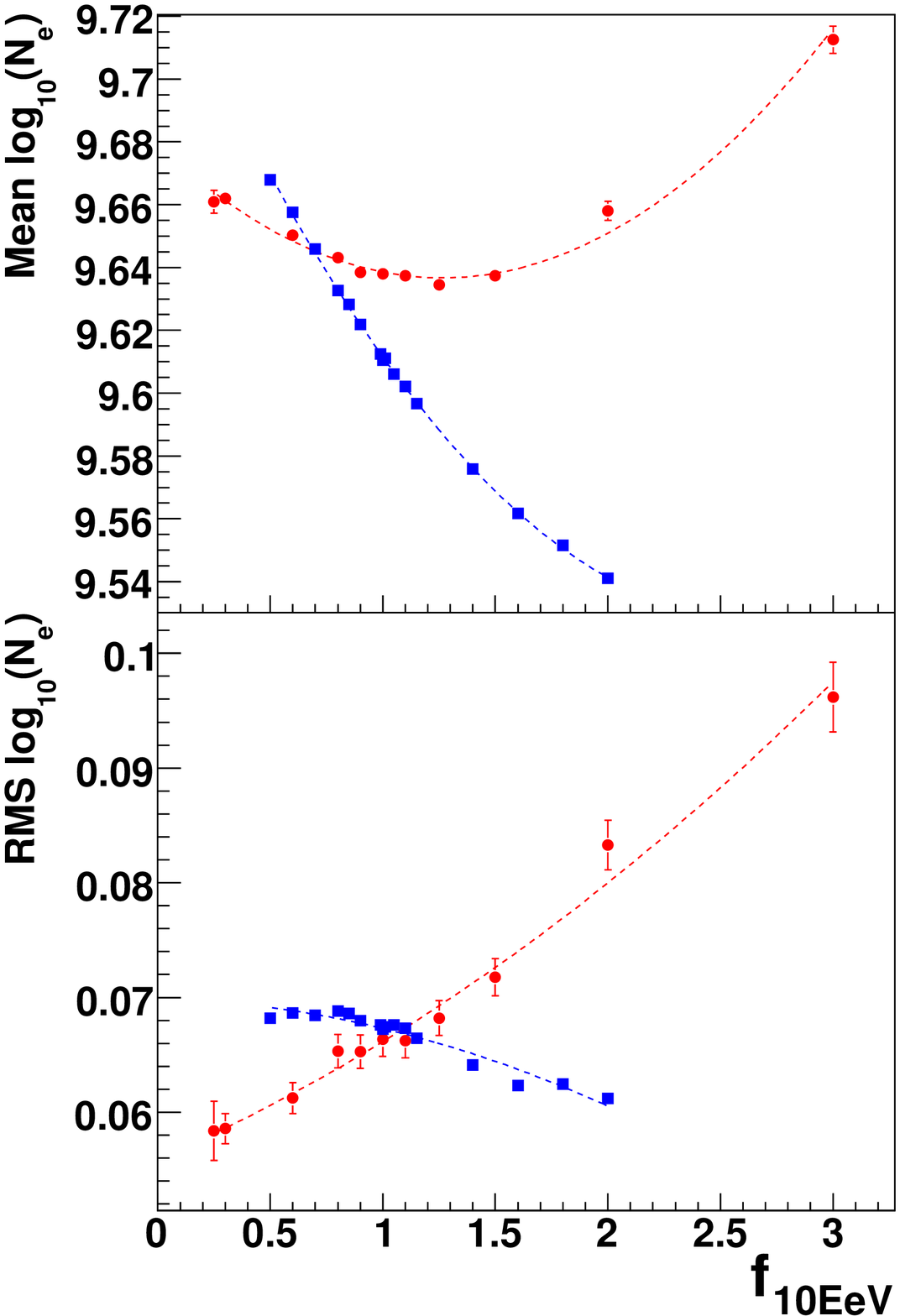}~
    \caption{Mean and RMS values for the resulting $X_{\rm max}$,
      $\left.N_{\mu}(X)\right|_{X=1000\textrm{~gcm}^{-2}}$ and 
      $\left.N_{\rm e}(X)\right|_{X=1000\textrm{~gcm}^{-2}}$
      distributions as a 
      function of $f_{\rm 10EeV}$ using 
      SIBYLL~2.1. For each data point with changed \emph{multiplicity}, 
      1000 air showers are simulated and 10000 for a changed
      \emph{cross section}. The dashed lines are polynomial fits of
      2$^{\rm nd}$ order to guide the eye.}
    \label{f:combined}
  \end{figure}
  To explore the impact of uncertainties of the present high energy
  hadronic interaction models on the interpretation of EAS observables, we
  modified the CONEX \cite{Bergmann:2006yz} program to change some of the interaction
  characteristics 
  during EAS simulation. To achieve this, individual hadronic interaction
  characteristics are altered by the energy-dependent factor  
  \begin{equation}
    f(E)=\left\{
    \begin{array}{l l}
      1 & \quad \mbox{\textit{E}$\le$1~PeV}\\
      1+(f_{\rm 10EeV}-1)\cdot
      \log_{10}(E/1\rm{PeV})/\log_{10}(10\rm{EeV}/1\rm{PeV} ) & \quad
      \mbox{\textit{E}$>$1~PeV}
    \end{array}\right.
  \end{equation}
  which was chosen to be $1$ below 1~PeV, because at these energies
  accelerator data is available (Tevatron corresponds to 1.8~PeV). Above
  1~PeV, $f(E)$ increases logarithmically with energy, reaching 
  the value of $f_{\rm 10EeV}$ at 10~EeV.\\ 
  The factor $f(E)$ is then used to re-scale
  specific characteristic properties of the high energy hadronic
  interactions such as the interaction cross section, secondary particle
  multiplicity or inelasticity.
  Obviously by doing this we may leave the parameter space
  allowed by the original model, but nevertheless one can get a clear
  impression of how the resulting EAS properties are depending on the specific
  interaction characteristics. \\
  We demonstrate the impact of a changing multiplicity $n_{\rm
    mult}$ and cross section $\sigma$ on the following, important air
  shower observables: shower maximum $X_{\rm max}$, and the total number of
  electrons $N_{\rm e}^{\rm rec}$, 
  as well as muons $N_{\mu}^{\rm rec}$ arriving at an observation level of $X_{\rm
  obs}=1000$~gcm$^{-2}$. Figure~\ref{f:mult} shows the range of
  extrapolations of $n_{\rm mult}$ used by the current hadronic interaction models and
  thus motivates the energy dependent re-scaling of $n_{\rm mult}$ by $0.3\ge f_{\rm
  10EeV}\ge3$. \\
  All simulations are performed for primary
  protons at 10~EeV using the SIBYLL~2.1\cite{Engel:1999db} interaction
  model. Figure~\ref{f:combined} summarizes the results, which are discussed
  below.\vspace*{.1cm}\\    
  {\bf Multiplicity of secondary particle production}\\
  The effect of a changed multiplicity on the $X_{\rm max}$-distribution is a
  shift to shallower $X_{\rm max}$ with 
  increasing $n_{\rm mult}$. This is what is already predicted by the extended
  Heitler model \cite{Matthews:2005sd}
  \begin{equation}
    \label{eqn:extendentHeitler}
    X_{\rm max} \propto \lambda_r \cdot \ln \frac{E_0}{n_{\rm
	mult}\cdot E^{\;\rm e.m.}_{\rm crit}},
  \end{equation}
  where $\lambda_r$ is the electromagnetic radiation length and $E^{\;\rm
  e.m.}_{\rm crit}$ the critical energy in air. 
  This is a consequence of the distribution of the same energy onto a growing
  number of particles. The 
  resulting lower energy electromagnetic sub-showers reach their maximum
  earlier. The impact on the RMS of the $X_{\rm max}$-distribution is small, but
  there is a trend to smaller fluctuations for an increasing number of
  secondaries. \\
  The total muon number after 1000~gcm$^{-2}$ of shower development is rising 
  if the multiplicity increases. This reflects the overall increased
  number of particles. The fluctuations are not significantly affected.\\ 
  More interesting is the impact on the electron number $N_{\rm e}^{\rm rec}$, which
  shows a minimum close to $f_{\rm 10EeV}=1$. The rising
  trend in the direction of smaller $n_{\rm 
    mult}$ can be explained by the increase of $X_{\rm max}$ and therefore the
  shower maximum coming closer to the observation level. On the other hand the
  rising trend in the direction of larger $n_{\rm mult}$ is again just the
  consequence of a generally growing number of particles. In contrary
  to the muon number the RMS does significantly change while $n_{\rm mult}$ gets
  larger. This can be explained by the strong dependence of fluctuations in
  $N_{\rm e}^{\rm rec}$ from the distance to the shower maximum.\vspace*{.1cm}\\
  {\bf Cross section}\\
  By construction, scaling the cross section does affect all hadronic
  interactions above 1~PeV, not only the first interaction. \\
  The mean as well as the RMS of the $X_{\rm max}$-distribution are decreasing
  with an increasing cross section. The effect is very pronounced, since the depth of 
  the first interaction $X_{\rm 1}$ is affected as well as the shower startup
  phase. Both effects are pointing to the same
  direction. This makes $X_{\rm max}$ a very sensitive observable for a cross
  section measurement.\\ 
  The impact on the muon number $N_{\mu}^{\rm rec}$ is not very large. Since the shower 
  maximum moves away from the observation level with increasing cross
  section, we just see the slow decrease of the muon number at late shower
  development stages, while the fluctuation of $N_{\mu}^{\rm rec}$ stay basically constant.\\
  The mean electron number as well as its fluctuations depend strongly on the 
  distance of $X_{\rm max}$ from the observation level. Combined with the
  influence of the modified 
  cross section on $X_{\rm max}$ this explains well the strong decrease of the
  mean $N_{\rm e}^{\rm rec}$ as well as the RMS with increasing cross section. At very small
  cross sections the shower maximum comes very close to the observation level,
  which can be observed as a flattening in the mean $N_{\rm e}^{\rm rec}$ and the decrease
  of the fluctuations in $N_{\rm e}^{\rm rec}$ against the trend of increasing
  fluctuations of the position of the shower maximum itself.
  
  \section{Summary}

  \begin{figure}[t!]
    \centering
    \includegraphics[width=.74\linewidth]{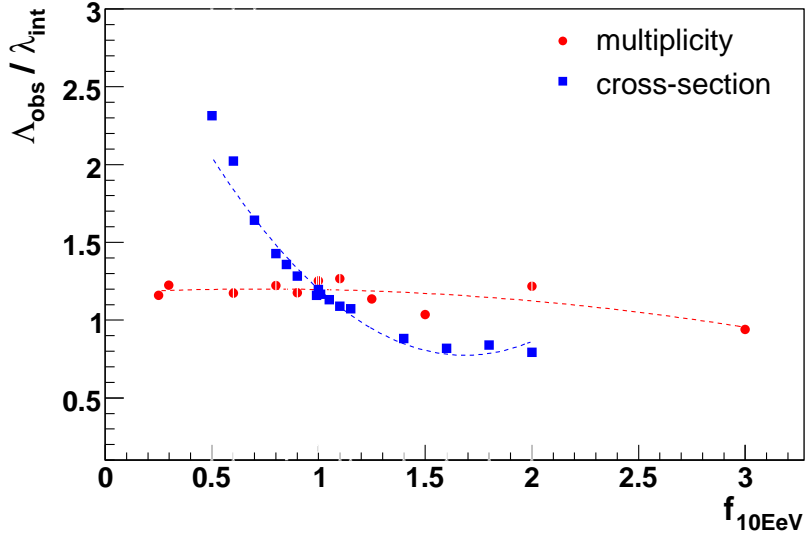}
    \caption{Simulated $k_{\rm X}$-factors ($k_{\rm X}=\Lambda^{\rm X}_{\rm
	obs}/\lambda_{\rm int}$) for SIBYLL at 10~EeV with modified cross section and
      multiplicity. $\Lambda_{\rm obs}^{\rm X}$ is obtained by an exponential
      fit to the tail of the resulting CONEX $X_{\rm max}$-distribution. An
      ideal detector is assumed, hence $k_{\rm X}=k_{\Delta X_{\rm 1}}$. The
	polynomial fits 
      of 2$^{\rm nd}$ order are only plotted to guide the eye.} 
    \label{f:KFactors}
  \end{figure}

  All methods of EAS-based cross section
  measurements are very similar and thus suffer from the same limitations.
  \begin{itemize}
    \item The values of all $k$-factors must be retrieved from massive
      Monte-Carlo 
      simulations. All analysis attempts so far have only calculated
      the combined factor of $k_{\rm S}$, respectively $k_{\rm X}$.
    \item $k$-factors depend on the resolution of the experiment and can
      therefore 
      not be transferred simply to other experiments.
    \item $k_{\rm X}$-factors are inherently different from $k_{\rm
      S}$-factors and can therefore not be transferred from an $X_{\rm
      max}$-tail analysis to that of ground based frequency attenuation or
      vice versa.  
    \item It cannot be disentangled whether a measurement of $\Lambda_{\rm
      obs}$ can be attributed to $\lambda_{\rm int}$ entirely or at least
      partly to 
      changed fluctuations in $\Delta X_{\rm 1}$ and/or $\Delta X_{\rm 2}$.
    \item Generally the $P_{\rm 1}$ and $P_{\rm 2}$ distributions have a
      complex shape and therefore the integrations of (\ref{eqn::frequency})
      and (\ref{eqn::XmaxTail}) to yield the approximations
      (\ref{eqn:FrequencyLambda}) and (\ref{eqn:XmaxTailLambda}) are 
      leading to non-exponential contributions.
    \item Any non-exponential contribution creates a strong dependence of the
      fitted $\Lambda_{\rm obs}$ on the chosen fitting range
      \cite{AlvarezMuniz:2004bx}. A strong non-exponential contribution makes
      the $k$-factor analysis unusable. 
    \item It can be shown that the $P_{\rm 1}(\Delta X_{\rm
      1})$-distributions is very sensitive to changes of the high energy
      hadronic interaction characteristics and thus $P(\Delta
      X)=f(\sigma,n_{\rm mult},...)$ is a 
      function of $\sigma$ , $n_{\rm mult}$ and other high energy model
      parameters. Consequently this also makes the $k$-factors
      depending on the high energy interaction characteristics 
      $k=f(\sigma,n_{\rm mult},...)$, which certainly must be
      considered for any cross section analysis.
  \end{itemize}  
  \vspace*{.3cm}
  \noindent
  In Fig.~\ref{f:KFactors} we show how the here presented simulations can be
  used to quantify the uncertainty caused in the $k$-factors due to the dependence  
  on $n_{\rm mult}$ to about $\pm\sim0.1$ for a variation of
  the multiplicity by a factor from 0.3 up to 3. It is clear that even without
  considering the multiplicity as a possible source of uncertainty the
  $\sigma$-dependence of the $k$-factors certainly needs to be taken into
  account. Otherwise a 
  systematic shift will be introduced into the resulting $\sigma_{\rm p-air}$,
  since part of the observed signal in $\Lambda_{\rm obs}$ is wrongly assigned
  to $\lambda_{\rm int}$, while in fact it must be attributed to
  $k(\sigma,n_{\rm mult},\dots)$ \cite{Ulrich:2007ij}. This has not been
  considered in any EAS-based $\sigma_{\rm p-air}$ measurement so far.\\ 
  
  \vspace*{-.5cm}
  \begin{footnotesize}
    \bibliographystyle{blois07} 
		      {\raggedright
			\bibliography{rulrich}

\providecommand{\etal}{et al.\xspace}
\providecommand{\href}[2]{#2}
\providecommand{\coll}{Coll.}
\catcode`\@=11
\def\@bibitem#1{%
\ifmc@bstsupport
  \mc@iftail{#1}%
    {;\newline\ignorespaces}%
    {\ifmc@first\else.\fi\orig@bibitem{#1}}
  \mc@firstfalse
\else
  \mc@iftail{#1}%
    {\ignorespaces}%
    {\orig@bibitem{#1}}%
\fi}%
\catcode`\@=12
\begin{mcbibliography}{10}

\bibitem{Grigorov:1965aa}
N.~L. Grigorov {\em et al.}~(1965).
\newblock Proc. of 9th Int. Cosmic Ray Conf. (London), vol.~1, p. 860\relax
\relax
\bibitem{Yodh:1972fv}
G.~B. Yodh, Y.~Pal, and J.~S. Trefil,
\newblock Phys. Rev. Lett.{} {\bf 28},~1005~(1972)\relax
\relax
\bibitem{Nam:1975aa}
R.~A. Nam, S.~I. Nikolsky, V.~P. Pavluchenko, A.~P. Chubenko, and V.~I.
  Yakovlev~(1975).
\newblock In Proc. of 14th Int. Cosmic Ray Conf. (Munich), vol.~7,
  p.~2258\relax
\relax
\bibitem{Siohan:1978zk}
F.~Siohan {\em et al.},
\newblock J. Phys.{} {\bf G4},~1169~(1978)\relax
\relax
\bibitem{Mielke94}
H.~H. Mielke, M.~F\"oller, J.~Engler, and J.~Knapp,
\newblock J. Phys. G{} {\bf 20},~637~(1994)\relax
\relax
\bibitem{Baltrusaitis:1984ka}
R.~M. Baltrusaitis {\em et al.},
\newblock Phys. Rev. Lett.{} {\bf 52},~1380~(1984)\relax
\relax
\bibitem{Honda:1993kv}
M.~Honda {\em et al.},
\newblock Phys. Rev. Lett.{} {\bf 70},~525~(1993)\relax
\relax
\bibitem{Aglietta:1999bd}
M.~Aglietta {\em et al.}~(1999).
\newblock In Proc. of 26th International Cosmic Ray Conference (ICRC 99), Salt
  Lake City, Utah, 17-25 Aug 1999, vol.~1, p.~143\relax
\relax
\bibitem{Hara:1983pa}
T.~Hara {\em et al.},
\newblock Phys. Rev. Lett.{} {\bf 50},~2058~(1983)\relax
\relax
\bibitem{Knurenko:1999cr}
S.~P. Knurenko, V.~R. Sleptsova, I.~E. Sleptsov, N.~N. Kalmykov, and S.~S.
  Ostapchenko~(1999).
\newblock In Proc. of 26th International Cosmic Ray Conference (ICRC 99), Salt
  Lake City, Utah, 17-25 Aug 1999, vol.~1, p.~372-375\relax
\relax
\bibitem{Belov:2006mb}
K.~Belov,
\newblock Nucl. Phys. Proc. Suppl.{} {\bf 151},~197~(2006)\relax
\relax
\bibitem{Ranft:1994fd}
J.~Ranft,
\newblock Phys. Rev.{} {\bf D51},~64~(1995)\relax
\relax
\bibitem{Pierog:2006qv}
T.~Pierog and K.~Werner~(2006).
\newblock \href{http://www.arXiv.org/abs/astro-ph/0611311}{{\tt
  astro-ph/0611311}}\relax
\relax
\bibitem{Drescher:2000ha}
H.~J. Drescher, M.~Hladik, S.~Ostapchenko, T.~Pierog, and K.~Werner,
\newblock Phys. Rept.{} {\bf 350},~93~(2001).
\newblock \href{http://www.arXiv.org/abs/hep-ph/0007198}{{\tt
  hep-ph/0007198}}\relax
\relax
\bibitem{Kalmykov:1997te}
N.~N. Kalmykov, S.~S. Ostapchenko, and A.~I. Pavlov,
\newblock Nucl. Phys. Proc. Suppl.{} {\bf 52B},~17~(1997)\relax
\relax
\bibitem{Ostapchenko:2005nj}
S.~Ostapchenko,
\newblock Phys. Rev.{} {\bf D74},~014026~(2006).
\newblock \href{http://www.arXiv.org/abs/hep-ph/0505259}{{\tt
  hep-ph/0505259}}\relax
\relax
\bibitem{Engel:1999db}
R.~Engel, T.~K. Gaisser, T.~Stanev, and P.~Lipari~(1999).
\newblock In Proc. of 26th International Cosmic Ray Conference (ICRC 99), Salt
  Lake City, Utah, 17-25 Aug 1999, p.~415-418\relax
\relax
\bibitem{Fletcher:1994bd}
R.~S. Fletcher, T.~K. Gaisser, P.~Lipari, and T.~Stanev,
\newblock Phys. Rev.{} {\bf D50},~5710~(1994)\relax
\relax
\bibitem{Ulrich:2006hb}
R.~Ulrich, J.~Blumer, R.~Engel, F.~Schussler, and M.~Unger~(2006).
\newblock In Proc. of XIV ISVHECRI 2006, Weihai, China, 2006,
  \href{http://www.arXiv.org/abs/astro-ph/0612205}{{\tt
  astro-ph/0612205}}\relax
\relax
\bibitem{Bergmann:2006yz}
T.~Bergmann {\em et al.},
\newblock Astropart. Phys.{} {\bf 26},~420~(2007).
\newblock \href{http://www.arXiv.org/abs/astro-ph/0606564}{{\tt
  astro-ph/0606564}}\relax
\relax
\bibitem{Matthews:2005sd}
J.~Matthews,
\newblock Astropart. Phys.{} {\bf 22},~387~(2005)\relax
\relax
\bibitem{AlvarezMuniz:2004bx}
J.~Alvarez-Muniz, R.~Engel, T.~K. Gaisser, J.~A. Ortiz, and T.~Stanev,
\newblock Phys. Rev.{} {\bf D69},~103003~(2004).
\newblock \href{http://www.arXiv.org/abs/astro-ph/0402092}{{\tt
  astro-ph/0402092}}\relax
\relax
\bibitem{Ulrich:2007ij}
R.~Ulrich, J.~Blumer, R.~Engel, F.~Schussler, and M.~Unger~(2007).
\newblock In Proc. of 30th International Cosmic Ray Conference (ICRC 07),
  Merida, Mexico, 2007, 2007, vol.~1, p.~143,
  \href{http://www.arXiv.org/abs/arXiv:0706.2086 [astro-ph]}{{\tt
  arXiv:0706.2086 [astro-ph]}}\relax
\relax
\end{mcbibliography}
		      }
  \end{footnotesize}
  
\end{document}